\documentclass{article}
\usepackage{amssymb}
\usepackage{amsmath}

\input{tcilatex}
\begin{document}

\title{Asymptotically Stationary and Static Space-times and Shear Free Null
Geodesic Congruences}
\author{T. M. Adamo$^{1}$, E. T. Newman$^{2}$ \\
$^{1}$Dept. of Mathematics, Univ. of Pittsburgh, \ Pgh. PA\\
$^{2}$Dept of Physics and Astronomy, Univ. of Pittsburgh, Pgh. PA}
\date{revised 6.11.09}
\maketitle

\begin{abstract}
In classical electromagnetic theory, one formally defines the \textit{%
complex dipole moment} (the electric plus 'i' magnetic dipole) and then
computes (and defines) the \textit{complex center of charge} by transforming
to a complex frame where the complex dipole moment vanishes. \ Analogously
in asymptotically flat space-times it has been shown that one can determine
the complex center of mass by transforming the complex gravitational dipole
(mass dipole plus 'i' angular momentum) (via an asymptotic tetrad
trasnformation) to a frame where the complex dipole vanishes.

\ We apply this procedure to such space-times which are asymptotically
stationary or static, and observe that the calculations can be performed
exactly, without any use of the approximation schemes which must be employed
in general. \ In particular, we are able to exactly calculate complex center
of mass and charge world-lines for such space-times, and - as a special case
- when these two complex world-lines coincide, we recover the Dirac value of
the gyromagnetic ratio.
\end{abstract}

\section{Introduction}

It is the purpose of this work to examine the special case of asymptotically 
\textit{stationary or static} space-times in the context of the recently
developed method for the identification of physical quantities from the
geometric quantities in general asymptotically flat space-times \cite%
{UCF,PhysicalContent,Review}. In the general case, due to the complexities
and non-linearities, approximations and restrictions on the spherical
harmonic expansions must be employed. In this special case, however, the
analysis can be done simply and exactly and is in full agreement with the
more general approximate work. \ In addition, it gives a clearer picture of
what the physical identification procedure is.

The principle stage on which this identification method is applied is the
future conformal boundary of the space-time, future null infinity (i.e.,
Penrose's $\mathfrak{I}^{+}$)$.$ $\mathfrak{I}^{+},$ which is a null (3-D)
surface, is coordinatized by the so-called Bondi coordinates $(u,\zeta ,%
\overline{\zeta })$ with $u=$constant cross-sections and ($\zeta ,\overline{%
\zeta }$) labeling its null generators (null geodesics). The asymptotically
flat Einstein equations in the spin-coefficient formalism are integrated in
the neighborhood of $\mathfrak{I}^{+}$ leading to the asymptotic behavior of
the spin-coefficient version of the Weyl tensor. \ These asymptotic Weyl
tensor components are objects that live on $\mathfrak{I}^{+}$ (i.e., are
functions only of $(u,\zeta ,\bar{\zeta})$)$.$ From a standard procedure 
\cite{PhysicalContent,Review}, a\ spherical harmonic term ($l=1$) of a
specific Weyl tensor component is identified with the complex dipole moment, 
$D_{\mathbb{C}}^{i}=(D_{(mass)}^{i}+ic^{-1}J^{i}),$ (mass dipole plus '$i$' $%
\times $ angular momentum).

The next step is to consider how the Weyl tensor, and hence $D_{\mathbb{C}%
}^{i},$ transforms under a change in the Bondi coordinates and tetrad (or a
generalization of this) at $\mathfrak{I}^{+}$ and thereby find the
transformation that produces the new $D_{\mathbb{C}}^{i}$ \cite%
{PhysicalContent,Review}$.$ Setting the new $D_{\mathbb{C}}^{i}=0$ defines
the \textit{complex center of mass} which turns out to be a complex
world-line in complex Minkowski space. \ Ordinarily, the calculations
required to transform this Weyl component to zero are quite complicated for
a general asymptotically flat space-time; indeed, exact computations are
usually impossible, and approximation schemes must be employed. \ In
practice, this is done in two ways: perturbation terms are expanded only to
second order, and harmonic expansions are truncated at the $l=2$
contributions (e.g., \cite{PhysicalContent,Review,AdamoNewman}). \ In the
present instance of asymptotically stationary (or static) space-times, the
calculations simplify and may be performed exactly.

A most important technical tool for these calculations comes from an
analysis of null geodesic congruences (NGCs) and specifically from the
regular shear-free or asymptotically shear-free NGCs \cite{Aronson,GCE1},
all of which can be constructed from solutions of the so-called Good Cut
equation \cite{Hspace}. The important point is the observation \cite%
{PhysicalContent,Review} that \textit{regular} asymptotically shear-free
null geodesic congruences in asymptotically flat space-times (or \textit{%
regular }shear-free congruences in Minkowski space-time) are determined by
the free choice of a complex world-line in an auxiliary complex Minkowski
space-time: the freedom in the choice of solutions to the Good Cut equation.
It turns out that setting the new $D_{\mathbb{C}}^{i}$ to zero uniquely
determines a specific world-line, referred to as the \textit{complex center
of mass. }This procedure is analogous (with a complex generalization) to the
situation in classical electromagnetic theory, where the center of charge
world-line is found by transforming to a system where the electric dipole
vanishes. \ The asymptotic Bianchi Identities \cite{NewmanTod} determine the
evolution equations for this line, which in turn allows for the physical
identifications.\ Furthermore, these results agree exactly with those from
already well known stationary (or static) space-times such as the Kerr
metric.

In this note we show how the asymptotic Weyl tensor can be found and the
asymptotic Banchi Identities easily integrated. \ From these results it is
simple to show from supertranslation in the asymptotic symmetry group (the
Bondi-Metzner-Sachs (BMS) group \cite{BMS}) that a Bondi coordinate/tetrad
system can be constructed so that the asymptotic (Bondi) shear vanishes.
This allows us to work with the \textit{homogeneous} Good Cut equation,
meaning that we have no need to force the termination of spherical harmonic
expansions. \ In turn, we are then able to compute exactly the asymptotic
tetrad transformation necessary to find the complex center of mass
world-line and allow us to identify and give a kinematic descriptions of the
mass, linear three-momentum and intrinsic spin of the system. \ 

For completeness, we include the Maxwell field in the discussion, i.e., we
are considering the asymptotically flat stationary Einstein-Maxwell equations

\section{Complex Center of Mass in an Asymptotically Stationary or Static
Space-time}

A space-time is stationary if it has a time-like Killing vector field (e.g.,
the Kerr metric) and static if the Killing field is surface forming (e.g.,
the Schwarzschild metric) \cite{Wald}. \ In both cases, the metric can be
written in coordinates that make it manifestly time-independent. \ For our
purposes, we define asymptotically stationary (or static) space-times to
mean that only all asymptotic variables of the space-time under
consideration are time-independent. \ Since we will utilize the well known
Bondi coordinate system $(u,r,\zeta ,\bar{\zeta})$, this simply means that
all asymptotic variables are $u$-independent. \ For instance, when
considering the Peeling property of the Weyl tensor in the spin coefficient
(SC) formalism \cite{NewmanTod},%
\begin{align}
\psi _{0}& =\psi _{0}^{0}r^{-5}+O(r^{-6}),  \label{peeling} \\
\psi _{1}& =\psi _{1}^{0}r^{-4}+O(r^{-5}),  \notag \\
\psi _{2}& =\psi _{2}^{0}r^{-3}+O(r^{-4}),  \notag \\
\psi _{3}& =\psi _{3}^{0}r^{-2}+O(r^{-3}),  \notag \\
\psi _{4}& =\psi _{4}^{0}r^{-1}+O(r^{-2}),  \notag
\end{align}%
it follows that%
\begin{equation}
\partial _{u}\psi _{k}^{0}\equiv \dot{\psi}_{k}^{0}=0,\ \ k=0,1,2,3,4,
\label{TI}
\end{equation}%
or that 
\begin{equation}
\psi _{k}^{0}=\psi _{k}^{0}(\zeta ,\bar{\zeta}).  \label{no u}
\end{equation}

The same applies to the asymptotic Maxwell field:%
\begin{eqnarray}
\phi _{0} &=&\frac{\phi _{0}^{0}}{r^{3}}+O(r^{-4}),  \label{Max1} \\
\phi _{1} &=&\frac{\phi _{1}^{0}}{r^{2}}+O(r^{-3}),  \notag \\
\phi _{2} &=&\frac{\phi _{2}^{0}}{r}+O(r^{-2}),  \notag \\
\phi _{k}^{0} &=&\phi _{k}^{0}(\zeta ,\bar{\zeta}).
\end{eqnarray}

Recall the asymptotic Bianchi identities for the Weyl and Maxwell tensors in
the SC formalism \cite{NewmanTod}:%
\begin{align}
\dot{\psi}_{2}^{0}& =-\eth \psi _{3}^{0}+\sigma ^{0}\psi _{4}^{0}+k\phi
_{2}^{0}\bar{\phi}_{2}^{0},  \label{ASB1} \\
\dot{\psi}_{1}^{0}& =-\eth \psi _{2}^{0}+2\sigma ^{0}\psi _{3}^{0}+2k\phi
_{1}^{0}\bar{\phi}_{2}^{0},  \label{ASB2} \\
\dot{\psi}_{0}^{0}& =-\eth \psi _{1}^{0}+3\sigma ^{0}\psi _{2}^{0}+3k\phi
_{0}^{0}\bar{\phi}_{2}^{0},  \label{ASB3} \\
\psi _{4}^{0}& =-\bar{\sigma}^{0\cdot \cdot },  \label{ASB4'} \\
\psi _{3}^{0}& =\eth \bar{\sigma}^{0\cdot },  \label{ASB5'} \\
\dot{\phi}_{1}^{0}& =-\eth \phi _{2}^{0},  \label{ASB4} \\
\dot{\phi}_{0}^{0}& =-\eth \phi _{1}^{0}+\sigma ^{0}\phi _{2}^{0},
\label{ASB5} \\
k& =2Gc^{-4},
\end{align}%
where $\sigma ^{0}$ is the complex Bondi shear (the free characteristic
data) for the space-time, coming from the expansion of the full shear:%
\begin{equation}
\sigma =\frac{\sigma ^{0}}{r^{2}}+O(r^{-4}),  \label{Shear}
\end{equation}%
and $\eth $ is the well known spin-weight operator on $S^{2}$. Additionally,
we have the Bondi Mass Aspect as%
\begin{equation}
\Psi \equiv \psi _{2}^{0}+\eth ^{2}\bar{\sigma}^{0}+\sigma ^{0}\bar{\sigma}%
^{0\cdot },  \label{MA1}
\end{equation}%
satisfying the familiar reality condition (derived from the SC equations):%
\begin{equation}
\Psi =\bar{\Psi}.  \label{MA2}
\end{equation}%
For the stationary (or static) case we have%
\begin{equation*}
\sigma ^{0}=\sigma ^{0}(\zeta ,\overline{\zeta }).
\end{equation*}

In an asymptotically stationary (or static) space-time, the Eqs.(\ref{ASB1})
- (\ref{ASB5}) reduce to:%
\begin{align}
0& =k\phi _{2}^{0}\bar{\phi}_{2}^{0}, \\
0& =-\eth \psi _{2}^{0}+2k\phi _{1}^{0}\bar{\phi}_{2}^{0}, \\
0& =-\eth \psi _{1}^{0}+3\sigma ^{0}\psi _{2}^{0}+3k\phi _{0}^{0}\bar{\phi}%
_{2}^{0}, \\
\psi _{4}^{0}& =0, \\
\psi _{3}^{0}& =0, \\
0& =-\eth \phi _{2}^{0}, \\
0& =-\eth \phi _{1}^{0}+\sigma ^{0}\phi _{2}^{0}, \\
\phi _{0}^{0}& =\phi _{0}^{0i}Y_{1i}^{1}(\zeta ,\overline{\zeta })+\cdots ,
\\
\phi _{0}^{0i}& =const,
\end{align}%
which immediately simplify to

\begin{eqnarray}
\phi _{2}^{0} &=&0,  \label{B1} \\
\eth \psi _{2}^{0} &=&0,  \label{B2} \\
\eth \psi _{1}^{0} &=&3\sigma ^{0}\psi _{2}^{0},  \label{B3} \\
\eth \phi _{1}^{0} &=&0,  \label{B4}
\end{eqnarray}%
while the mass aspect becomes:%
\begin{equation}
\Psi =\psi _{2}^{0}+\eth ^{2}\bar{\sigma}^{0}.  \label{MA3}
\end{equation}

Since $\psi _{2}$ is a spin-weight zero quantity, Eq.(\ref{B2}) tells us
immediately that $\psi _{2}^{0}$ contains only the $l=0$ harmonic in a
spherical harmonic expansion. \ Keeping this in mind, if we apply the
reality condition of (\ref{MA2}) to (\ref{MA3}), we obtain:%
\begin{equation}
\psi _{2}^{0}+\eth ^{2}\bar{\sigma}^{0}=\bar{\psi}_{2}^{0}+\bar{\eth }%
^{2}\sigma ^{0},  \label{Real}
\end{equation}%
or%
\begin{equation}
\func{Im}(\psi _{2}^{0})=-\frac{i}{2}\left( \bar{\eth }^{2}\sigma ^{0}-\eth
^{2}\bar{\sigma}^{0}\right) =\func{Im}(\bar{\eth }^{2}\sigma ^{0}).
\label{Im}
\end{equation}%
The imaginary part of $\psi _{2}^{0}$ is thus determined by the "magnetic"
part of the asymptotic shear. Since $\sigma ^{0}$ has spin-weight $2$ and
contains only $l\geq 2$ harmonics in its harmonic expansion, it follows that
that the magnetic portion of $\sigma ^{0}$ must vanish since we already
established that $\psi _{2}^{0}$ contains only $l=0$ harmonic contributions.

Now, under the operation of a supertranslation subgroup of the BMS group 
\cite{BMS}, the Bondi time coordinate transforms as:%
\begin{equation}
u\rightarrow u^{\prime }=u+\alpha (\zeta ,\overline{\zeta }),  \label{St}
\end{equation}%
for an arbitrary analytic function on the two-sphere. \ By the Sachs
Theorem, the asymptotic shear transforms under such a supertranslation
according to \cite{NewmanTod}:%
\begin{equation}
\sigma ^{0}\rightarrow \sigma ^{0\prime }=\sigma ^{0}+\eth ^{2}\alpha (\zeta
,\overline{\zeta }),  \label{ST}
\end{equation}%
so by the proper choice of the function $\alpha (\zeta ,\overline{\zeta })$,
the real ("electric") portion of the asymptotic shear may be set equal to
zero. \ Hence, we can make both the "electric" and the "magnetic" parts of
the asymptotic shear to vanish, and in an appropriate Bondi frame, we have
that 
\begin{equation}
\sigma ^{0}=0.  \label{sigma=0}
\end{equation}

Using the vanishing of $\sigma ^{0}$ we have from Eqs.(\ref{B2})-(\ref{B4})
that $\psi _{2}^{0}$ and $\phi _{1}^{0}$ have only an $l=0$ part, while $%
\psi _{1}^{0}$ only has an $l=1$ part. They turn out to be proportional to
the mass, the charge and the complex gravitational dipole moment
respectively.

Before turning to the recently developed procedure for the identification of
the physical variables hidden in the geometry, we summarize our situation.
In a preferred Bondi system we have the geometric results (using standard\
canonical Bondi identifications) for the major terms \cite{Review}:

\begin{eqnarray}
\Psi &=&\psi _{2}^{0}=\Psi ^{0}+\Psi ^{i}Y_{1i}^{0}+\cdots  \label{Id1} \\
\Psi ^{0} &=&\psi _{2}^{0}=-\frac{2\sqrt{2}G}{c^{2}}M_{B}  \label{Id1*} \\
\Psi ^{i} &=&-\frac{6G}{c^{3}}P^{i}=0  \label{Id1**} \\
\psi _{1}^{0} &=&-\frac{6\sqrt{2}G}{c^{2}}%
(D_{(mass)}^{i}+ic^{-1}J^{i})Y_{1i}^{1},  \label{Id2} \\
\psi _{0}^{0} &=&Q_{\mathbb{C}}^{ij}Y_{2ij}^{2}+...,  \label{Id3} \\
\phi _{1}^{0} &=&q,  \label{Id4} \\
\phi _{0}^{0} &=&2(D_{\mathbf{E}}^{i}+iD_{\mathbf{M}}^{i})Y_{1i}^{1},
\label{Id4*} \\
\sigma ^{0} &=&0,  \notag
\end{eqnarray}%
where ($M_{B},q,D_{\mathbf{E}}^{i},D_{\mathbf{M}%
}^{i},D_{(mass)}^{i},J^{i},Q_{\mathbb{C}}^{ij}$) are, respectively: the
Bondi mass, the Coulomb charge, the (stationary) electric, magnetic and mass
dipoles, angular momentum and complex quadrupole (mass and spin) moments.

The recently developed identification procedure \cite{PhysicalContent,Review}
begins with two related ideas: (1) In flat space Maxwell theory one can
transform the origin of coordinates to the center of charge so that the
electric dipole moment associated with this origin vanishes. This suggests
that the same be done in general relativity (GR) to the center of mass so
that the mass dipole vanishes. (2) It raises the question: can one
generalize this transformation in GR and relate it to a remarkable property
of regular shear free or asymptotically shear free null geodesic congruences
(NGCs): that each such congruence is determined uniquely by a complex
world-line in an auxiliary complex Minkowski space \cite{GCE1,Review}?

The answer is yes. For general asymptotically flat space-times \cite{Twist}
one can chose, in the neighborhood of $\mathfrak{I}^{+},$ a family of (in
general twisting) null geodesics, i.e., a NGC, that `appears' to come from a
complex world-line in complex Minkowski space \cite{PhysicalContent}. By
appropriate choice of the congruence followed by rotating the Bondi tetrad
to a tetrad based on these new null geodesics, one can set the mass dipole
to zero. As an added bonus, since the world-line is complex it can be chosen
so that the angular momentum can be made to vanish. It is this general
procedure that we apply to the special case of asymptotically stationary
space-times.

In general shear-free or asymptotically shear-free NGCs are generated by the
solutions 
\begin{equation*}
u=G(\tau ,\zeta ,\bar{\zeta})
\end{equation*}%
of the so-call good cut equation:%
\begin{equation*}
\eth ^{2}G=\sigma ^{0}(G,\zeta ,\bar{\zeta}),
\end{equation*}%
which in our case of vanishing Bondi shear becomes the homogeneous good
equation,

\begin{equation}
\eth ^{2}G=0.  \label{GCE}
\end{equation}%
Its general regular solution is given by:%
\begin{eqnarray}
u &=&G(\tau ,\zeta ,\bar{\zeta})  \label{GCF} \\
&=&\xi ^{a}(\tau )l_{a}(\zeta ,\bar{\zeta})  \notag \\
&=&\frac{\sqrt{2}}{2}\xi ^{0}(\tau )-\frac{1}{2}\xi ^{i}(\tau
)Y_{1i}^{0}(\zeta ,\overline{\zeta }),  \notag
\end{eqnarray}%
with

\begin{eqnarray}
\hat{l}_{a}(\zeta ,\bar{\zeta}) &=&\frac{\sqrt{2}}{2(1+\zeta \bar{\zeta})}%
\left( 1+\zeta \bar{\zeta},-(\zeta +\bar{\zeta}),\ i(\zeta -\bar{\zeta}%
),1-\zeta \bar{\zeta}\right) ,  \label{l} \\
&=&\left( \frac{\sqrt{2}}{2},\frac{1}{2}Y_{1i}^{0}(\zeta ,\overline{\zeta }%
)\right) ,  \notag
\end{eqnarray}%
where $z^{a}=\xi ^{a}(\tau )$ is the complex Minkowski space world-line
parameterized by the complex $\tau .$ The relation given by Eq.(\ref{GCF})
describes a one-parameter family of complex cuts on the complexified $%
\mathfrak{I}^{+}.$ Though we are interested in the real values of $u,$ the
local complexification is essential. The reason lies in the following: the
asymptotically shear-free NGC is determined from Eq.(\ref{GCF}) by taking
its derivative and defining the stereographic angle field on $\mathfrak{I}%
^{+}$ by the parametric relations

\begin{eqnarray}
L(u,\zeta ,\bar{\zeta}) &=&\eth _{(\tau )}G(\tau ,\zeta ,\bar{\zeta})|_{\tau
=T{\small (}u,\zeta ,\bar{\zeta}{\small )}}  \label{L} \\
&=&\xi ^{a}(\tau )|_{\tau =T{\small (}u,\zeta ,\bar{\zeta}{\small )}%
}m_{a}(\zeta ,\bar{\zeta})  \notag \\
&=&\xi ^{i}(\tau )|_{\tau =T{\small (}u,\zeta ,\bar{\zeta}{\small )}%
}Y_{1i}^{1}(\zeta ,\bar{\zeta}),  \notag
\end{eqnarray}%
where $\tau =T(u,\zeta ,\bar{\zeta})$ is the inverse function to Eq.(\ref%
{GCF}) and $\eth _{(\tau )}$ indicates the application of the $\eth $%
-operator while the variable $\tau $ is held constant. The real values of $u$
must be used \textit{only after} the action of the derivative.

The $L(u,\zeta ,\bar{\zeta})$ constructed in this manner determines a null
direction field at $\mathfrak{I}^{+}$ that in turn determines an
asymptotically shear-free NGC. \ In addition it plays a key role in the
physical identifications.

We now single out a particular world-line $\xi ^{a}$ which is to represent a
"complex center of mass" for the system, thus choosing a particular good cut
function, Eq.(\ref{GCF}). \ To do this, we recall (\ref{Id2}) that the $l=1$
contribution from $\psi _{1}^{0}$ is identified as the complex gravitational
dipole \cite{Review}:%
\begin{equation}
\psi _{1}^{0i}=-\frac{6\sqrt{2}G}{c^{2}}D_{\mathbb{C}}^{i}=-\frac{6\sqrt{2}G%
}{c^{2}}(D_{(mass)}^{i}+ic^{-1}J^{i}),  \label{Dip}
\end{equation}%
where $D_{(mass)}^{i}$ is the mass dipole and $J^{i}$ is the angular
momentum (differences with other angular momentum identification theories
should vanish in this case, due to the shear-freeness and dynamical
simplicity). \ Hence, in a center of mass frame, we expect $\psi
_{1}^{0i\ast }=0$. \ The transformation of the Bondi tetrad to the tetrad
associated with the shear-free congruence is given asymptotically by the
null rotation \cite{NewmanTod}:%
\begin{align}
l^{a}& \rightarrow l^{a\ast }=l^{a}-\frac{\bar{L}}{r}m^{a}-\frac{L}{r}\bar{m}%
^{a}+O(r^{-2}),  \label{NullRot} \\
m^{a}& \rightarrow m^{a\ast }=m^{a}-\frac{L}{r}n^{a}+O(r^{-2}),  \notag \\
n^{a}& \rightarrow n^{a\ast }=n^{a},  \notag
\end{align}%
where $L$ is a holomorphic, spin-weight one function given by Eq.(\ref{L})
and $\{l^{a},n^{a},m^{a},\bar{m}^{a}\}$ is the well-known Bondi null tetrad,
chosen in this case so that $l^{a}$ is tangent to geodesics of the constant $%
u$ null hypersurfaces in the space-time. \ Under such a rotation, $\psi
_{1}^{0i}$ transforms, for the asymptotically stationary (or static) case,
as:%
\begin{equation}
\psi _{1}^{0i}\rightarrow \psi _{1}^{0i\ast }=\psi _{1}^{0i}-3L\psi
_{2}^{0}|^{i},  \label{NT2}
\end{equation}

Setting \ $\psi _{1}^{0i\ast }=0,$ then using $\psi _{1}^{0i}=-\frac{6\sqrt{2%
}G}{c^{2}}D_{\mathbb{C}}^{i}$ , $\psi _{2}^{0}=-\frac{2\sqrt{2}G}{c^{2}}%
M_{B} $ and $L(\tau ,\zeta ,\bar{\zeta})=\xi ^{i}(\tau )Y_{1i}^{1}(\zeta ,%
\bar{\zeta}),$ Eq.(\ref{NT2}) immediately leads to%
\begin{equation}
\psi _{1}^{0i}=-\frac{6\sqrt{2}G}{c^{2}}M_{B}\xi ^{i},  \label{p1}
\end{equation}%
or, the kinematic expression for the complex dipole,%
\begin{eqnarray}
D_{\mathbb{C}}^{i} &=&M_{B}\xi ^{i},  \label{p2} \\
D_{(mass)}^{i} &=&M_{B}\xi _{R}^{i}, \\
c^{-1}J^{i} &=&M_{B}\xi _{I}^{i}.
\end{eqnarray}

Now, by the time-independence of the system (i.e., $P^{i}=0$), it follows
that the "spatial" part of our world-line, $\xi ^{i}$, must be a constant
3-vector. \ Although as it is written now, $\xi ^{i}$ is a function of $\tau 
$, we can invert (\ref{GCF}) to obtain the world-line as a function of $u$;
thus $u$-independence is equivalent to $\tau $-independence. \ For such a
constant vector, a Poincar\'{e} translation can be chosen to set the real
part, $\xi _{R}^{i}$, equal to zero, and an ordinary rotation can be made to
set the imaginary part to%
\begin{equation}
\xi _{I}^{i}=(0,0,\xi _{I}^{3}).  \label{WL1}
\end{equation}%
Thus, if we also demand that the tangent vector $v^{a}=\partial _{\tau }\xi
^{a}$ be normalized to length one, we obtain the center of mass world-line
as:%
\begin{equation}
\xi ^{a}=(\tau ,0,0,i\xi _{I}^{3}).  \label{WL2}
\end{equation}%
From equations (\ref{GCF}) and (\ref{L}) we thus have that%
\begin{eqnarray}
u &=&G(\tau ,\zeta ,\bar{\zeta})=\frac{\tau }{\sqrt{2}}-\frac{i}{2}\xi
_{I}^{3}Y_{1,3}^{0},  \label{Funcs} \\
L(\tau ,\zeta ,\bar{\zeta}) &=&i\xi _{I}^{3}Y_{1,3}^{1}.  \notag
\end{eqnarray}

Finally, we flesh out our physical identification via equation (\ref{Dip})
to get:%
\begin{eqnarray}
D_{(mass)}^{i} &=&0,  \label{R1} \\
J^{i} &=&S^{i}=cM_{B}\xi _{I}^{3}\delta _{3}^{i}=cM_{B}\xi _{I}^{i},
\label{R2}
\end{eqnarray}%
where we have made the conventional identification of the intrinsic spin as $%
S^{i}=$ $cM_{B}\xi _{I}^{i}$.

We note that there was no discussion of the higher stationary moments that
would appear in the higher powers of $r^{-1}$ in the expansion for the Weyl
component $\psi _{0}=\psi _{0}^{0}r^{-5}+\psi _{0}^{1}r^{-6}+\cdots .$They
are indeed there, but play no role in the transformation to the complex
center of mass. If they happen to vanish we are left with the asymptotic
Kerr or Kerr-Newman metrics

Note also that the Maxwell field also played no role in the discussion, due
to (\ref{B1}). However, there is a parallel discussion, given in the
following section, for the Maxwell field when one can transform to the
complex center of charge where both the electric and magnetic dipoles vanish.

\section{Center of Charge World-line}

From the asymptotic Maxwell equations we obtained, Eqs.(\ref{B1}), (\ref{Id4}%
) and (\ref{Id4*}),%
\begin{eqnarray}
\phi _{2}^{0} &=&0  \label{B1*} \\
\phi _{1}^{0} &=&q  \label{Id**} \\
\phi _{0}^{0} &=&2(D_{\mathbf{E}}^{i}+iD_{\mathbf{M}}^{i})Y_{1i}^{1},
\label{Id4***}
\end{eqnarray}

The higher multipole moments, hidden in the higher $r^{-n}$ terms in the
expansion of $\phi _{0}$ (see Eq.(\ref{Max1})), are not needed or used in
this discussion.

We now transform via the asymptotic null rotation, Eq.(\ref{NullRot}), to
the complex center of charge world-line where both the electric and magnetic
dipoles vanish. The $L$ and its associated complex world-line used$\ $in
this section, i.e., $L=\eta ^{i}Y_{1i}^{1},$ is in general independent of
the $L$ used in the previous section, $L=\xi ^{i}Y_{1i}^{1}.$

Under this rotation, Eq.(\ref{NullRot}), the Maxwell tensor component $\phi
_{0}^{0}$ transforms as:%
\begin{equation}
\phi _{0}^{0}\rightarrow \phi _{0}^{0\ast }=\phi _{0}^{0}-2L\phi
_{1}^{0}+L^{2}\phi _{2}^{0}.  \label{phi*}
\end{equation}

Using Eqs.(\ref{B1*})-(\ref{Id4***}) and (\ref{L}), with the assumption that 
$\phi _{0}^{0\ast }=0$ at the center of charge, we obtain from (\ref{phi*})
that%
\begin{equation}
q\eta ^{i}=D_{\mathbb{C}}^{i}=D_{\mathbf{E}}^{i}+iD_{\mathbf{M}}^{i}.
\label{CofCharge}
\end{equation}%
In analogy to the complex center of mass line, this determines the complex
center of charge world-line, 
\begin{equation*}
z^{a}=\eta ^{a}=(\tau ,q^{-1}D_{\mathbb{C}}^{i}),
\end{equation*}%
a straight time-like world-line displaced into the complex.

Note that this is just a complex generalization of finding the real center
of charge in electrostatics.

Since we have already fixed the Bondi system via the complex center of mass
argument we can not further simplify the Maxwell field. \ We however can
consider the very special case where the complex center of charge coincides
with the complex center of mass, $\eta ^{a}=\xi ^{a}.$ This leads to the
real relation

\begin{equation*}
q^{-1}D_{\mathbf{M}}^{i}=c^{-1}M_{B}^{-1}S^{i},
\end{equation*}%
or

\begin{equation}
\frac{D_{\mathbf{M}}^{i}}{S^{i}}=\frac{q}{cM_{B}},  \label{GMR}
\end{equation}%
which leads to the Dirac value of the gyromagnetic ratio, $g=2$.

\section{Conclusion}

As we can see from equations (\ref{WL2}), (\ref{R1}), and (\ref{R2}), the
case of an asymptotically stationary (or static) space-time provides us with
a very nice example for the application of the recently developed physical
identification theory based on the null rotation (\ref{NullRot}) to an
asymptotically shear-free NGC. \ In particular, the complex center of mass
world-line, Bondi mass and linear momentum, gravitational dipole, and
spin-angular momentum were all calculated quite easily, and without the need
to impose constraints on the harmonic expansions used or the order of the
expressions in terms of the world-line. \ Additionally, we found that these
results concurred exactly with what was to be expected from a stationary or
static space-time \textit{a priori}.

In particular, we found that such space-times have vanishing Bondi
three-momentum (as expected) and that all of their angular momentum is
intrinsic, or in other words, all angular momentum takes the form of
intrinsic spin. \ Furthermore, the spin expression obtained in (\ref{R2}) is
exactly that of the Kerr solution, probably the most important example of a
stationary space-time \cite{Kerr}. \ Our results represent the most general
asymptotic results for the stationary or static cases and are unaffected by
higher multipole considerations. \ Of course, for a static space-time, there
should be no spin at all, and this would correspond to a world-line whose
spatial part is real (i.e., $\xi _{I}^{i}=0$). \ When the two world-lines,
complex centers of mass and charge, coincide, we have a 'Dirac-like'
particle with $g=2$ from equation (\ref{GMR}).

\end{document}